\documentclass[twocolumn]{article}
\usepackage{pkuarticle}

\AtBeginDocument{%
  }

\usepackage{array}
\usepackage{algorithm}
\usepackage{algpseudocode}
\usepackage{amsmath}
\usepackage{booktabs}
\usepackage{multirow}
\usepackage{subcaption}
\usepackage{xspace}
\usepackage{tabularx}
\usepackage{tabulary}
\usepackage{enumitem}
\usepackage{lipsum}
\usepackage[normalem]{ulem}
\usepackage[numbers,sort&compress]{natbib}

\renewcommand{\dj}[1]{\textcolor{blue}{\textbf{[DJ: #1]}}}

\newcommand{\codename}{MCHA\xspace}
\renewcommand{\dj}[1]{\textcolor{black}{#1}}

\newcommand{\copyrightnotice}{%
  \begingroup
  \renewcommand{\thefootnote}{}%
  \footnotetext{\footnotesize
    This paper has been accepted for publication at the 59th IEEE/ACM
    International Symposium on Microarchitecture (MICRO). \copyright\ IEEE.
    Personal use of this material is permitted. Permission from IEEE must be
    obtained for all other uses, in any current or future media, including
    reprinting/republishing this material for advertising or promotional
    purposes, creating new collective works, for resale or redistribution to
    servers or lists, or reuse of any copyrighted component of this work in
    other works.}%
  \endgroup
}

\begin{document}

\renewcommand{\and}{,\ }
\title{MCHA: A \underline{M}emory-\underline{C}entric \underline{H}ierarchical \underline{A}rchitecture for Parallel-Sequential Computing}
\author{Daijing Shi \and Hongxiao Zhao \and Yihan Fu \and Zhan Chen \and Jiayi Li \and Yihang Zhu \and Anjunyi Fan \and Yaoyu Tao \and Yuchao Yang \and Bonan Yan}
\affiliation{Peking University, Beijing 100871, China\\
\small\texttt{bonanyan@pku.edu.cn}}




\maketitle
\copyrightnotice

\begin{abstract}
Emerging workloads, such as Multi-Agent Reinforcement Learning (MARL), large-scale neuromorphic computing, and probabilistic graphical models, intrinsically exhibit parallel-sequential computing patterns. While these tasks demand massive parallelism to achieve high throughput, they are severely bottlenecked by irregular data access patterns centralized to main memory. Consequently, conventional architectures face fundamental limitations when executing these workloads, primarily manifesting as global buffer saturation and memory-bound bottlenecks. To address these challenges, we propose the Memory-Centric Hierarchical Architecture (MCHA), a reconfigurable hardware solution tailored for parallel-sequential execution. MCHA leverages a hierarchical communication strategy that facilitates distributed, inter-core data routing, thereby significantly reducing the bandwidth burden on the global memory. Complementing the hardware, MCHA introduces a novel parallel-sequential programming model that utilizes event-driven conditional triggers to effectively hide data transmission latency within the execution pipeline. We benchmark MCHA against a diverse suite of parallel-sequential tasks, including MARL, motor variable control, and Markov random fields. Validated through our open-source, cycle-accurate simulator, MCHA demonstrates performance speedups ranging from 153.06$\times$ to 2456.96$\times$ over NVIDIA A100 GPUs on MARL workloads, while maintaining robust programming flexibility across other application domains. Furthermore, the architecture successfully reduces main memory access from 96\% to 5.44\%. When synthesized in a 28 nm process, the \codename implementation occupies an area footprint of 2.92mm$^2$ and consumes 115.36 mW of power at 200 MHz.
\codename is open-sourced at \url{https://github.com/carabdis/MCHA}.

\end{abstract}


\section{Introduction}\label{sec:intro}

\begin{figure}[t!]
    \centering
    \includegraphics[width=1\columnwidth]{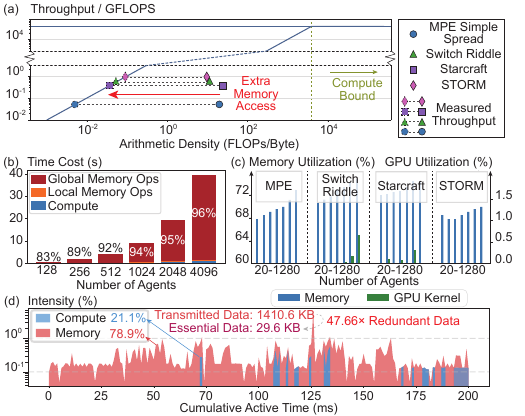}
    \caption{
    (a) Roofline model of the NVIDIA RTX 3090 with PSC applications;
        \dj{(b) Simulated access operation breakdown of the GPU architecture;
        (c) Measured memory account and computing intensity;}
        (d) Measured memory access and computing ratio during the active time.
    }
    \label{fig:1}
\end{figure}

Parallel-Sequential Computing (PSC) is a hybrid paradigm that combines concurrent execution of multiple tasks with stepwise sequential operations.
This approach seeks to balance computational efficiency, system scalability, and adherence to logical dependencies within a problem.
In contrast to purely parallel architectures (e.g., Tensor Processing Unit~\cite{jouppi2017datacenter} and the Loihi family~\cite{davies2018loihi}) and strictly sequential processing, PSC decomposes complex problems into independent subproblems and interdependent phases. This allows it to leverage distributed hardware while preserving data consistency, control flow, and algorithmic correctness.
Such a design mitigates key limitations of fully parallel execution, including synchronization overhead, data races, and bottlenecks described by Amdahl’s law~\cite{amdahl1967validity}. Simultaneously, it avoids the inefficiency of purely sequential processing when applied to large-scale or computationally intensive workloads.
By dynamically allocating parallelizable segments and enforcing sequential ordering where dependencies exist, the PSC execution model optimizes resource utilization and accelerates computation for data-intensive and algorithmically complex tasks. Consequently, it is widely adopted in high-performance computing, scientific simulation, and related domains.

PSC encompasses a broad spectrum of computing tasks, including Multi-Agent Reinforcement Learning (MARL)~\cite{chen2025review, gao2024hgpcn, hao2023exploration, nguyen2020deep, ning2024survey}, large-scale neuromorphic computing~\cite{beniaguev2021single, mead2002neuromorphic, roy2019towards, zhang2020system}, and probabilistic graphical models~\cite{li2024neural, kollovieh2024expected, arya2025sine}. Additionally, the Bulk Synchronous Parallel (BSP) paradigm, which underpins industrial data-intensive frameworks such as MapReduce~\cite{dean2008mapreduce}, Pregel~\cite{male2010pregel}, Spark~\cite{matei2012spark}, and PolyGraph~\cite{dadu2021polygraph}, is also integral to PSC.
These tasks demand massive parallelism to manage frequent, fine-grained interactions among thousands of entities (e.g. agents in MARL), with strict synchronization required at each time-step boundary.
In conventional compute-centric architectures (CPUs \& GPUs), these repeated interactions induce irregular data access patterns that saturate centralized main memory. Consequently, as depicted in Figure~\ref{fig:1}(a), deploying such workloads on GPUs or many-core CPUs results in execution throttled by the global memory bandwidth (i.e. the ``memory wall''~\cite{asanovic2006landscape, rogers2009scaling, ivanov2021data}).

The memory wall is exacerbated by the mismatch between irregular working sets and rigid hardware hierarchies. When on-chip storage fails to capture unpredictable access patterns, the system incurs frequent, high-latency DRAM fetches, leading to severe execution stalls~\cite{zhang2020system, burtscher2012quantitative, shin2018scheduling}. Our analysis of a grid-world MARL task on an NVIDIA RTX 3090 (Figure~\ref{fig:1}(b)) reveals a significant inefficiency: \textbf{data movement overhead accounts for 83\% to 96\% of total processing operations}, depending on the simulation scale. Figures~\ref{fig:1}(c) and (d) further demonstrate that the majority of execution cycles and energy are consumed by redundant memory traffic rather than productive computation.

Even recent reconfigurable hardware with data parallelism, such as Plasticine~\cite{prabhakar2017plasticine, prabhakar2024sambanova} or Leviathan~\cite{schwedock2024leviathan}, remains tethered to data-centric or compute-centric paradigms. Plasticine employs a coarse-grained reconfigurable architecture (CGRA) with dataflow compilation~\cite{chin2017cgra, li2025enhancing, qin2025picachu} to decouple memory access from task processing. However, the lack of transparency between memory and processing units still necessitates redundant data fetches. Leviathan~\cite{schwedock2024leviathan} utilizes in-cache near-data computing~\cite{aga2017compute, agarwal1995alewife, ainsworth2016graph, ainsworth2018event} but remains burdened by frequent cache coherence management overhead for PSC tasks.
Other Processing-In-Memory (PIM) or BSP domain-specific accelerators (DSAs) are either highly application-specific~\cite{chen2025heat, saed2025rayn, jang2025acce, kim2025pimba} or bottlenecked by global buffer (GB) access for synchronization~\cite{an2024stream, kim2023moca, kim2025pimcca, son2025pimnet, yu2025compass}. Consequently, existing works suffer from a ping-pong effect in the sequential part of PSC, where data is repeatedly transmitted between main memory and multi-level caches to satisfy rigid cache coherence protocols~\cite{lee2025beyond, ji2025re, liu2025system, sun2025m5}.

To address this challenge, we introduce \textbf{\codename}, a reconfigurable and memory-centric multi-chip computing architecture. By devising a multi-tiered inter-core communication network and a data-driven programming model, \codename effectively overcomes the data-transfer bottlenecks that limit the performance of GPUs, FPGAs, and CGRAs across a broad range of PSC tasks.
Our key contributions are summarized as follows:

\begin{itemize}[leftmargin=*, label*=$\diamond$]
\item \textbf{\codename Architecture}: After quantifying the inefficiency of traditional architectures for PSC workloads, we propose a novel reconfigurable hardware architecture. It replaces centralized global buffers with a distributed memory fabric interconnected via a Network-on-Chip (NoC), substantially lowering cache coherence overhead.

\item \textbf{Data-Driven Programming Model}: We develop a formal programming model and an associated workflow to efficiently map PSC tasks onto the proposed reconfigurable hardware platform.

\item \textbf{Comprehensive Architectural Analysis Framework}: We release an open-source, cycle-accurate simulator (available at \url{https://github.com/carabdis/MCHA}) to evaluate the throughput and scalability of \codename under various multi-chip configurations. Through extensive experiments, we derive new insights into hierarchical memory bandwidth limitations and propose effective mitigation strategies.
\end{itemize}

For evaluation, we implement \codename at the register-transfer level (RTL) using the TSMC 28nm HPC+ process design kit (PDK). Experiments across a suite of MARL benchmarks~\cite{terry2021pettingzoo, flair2024jaxmarl, foerster2016learning, lu2022model, foerster2018learning} show that a 4-chip \codename system delivers a speedup of 153.06$\times$ to 2456.96$\times$ compared to an NVIDIA A100 GPU. We further validate the generality of \codename using Motor Variable Control and BSP applications, where a 32-chip configuration achieves speedups of 1.17$\times$ and 3.90$\times$ over state-of-the-art DSAs. These results underscore the efficiency of \codename in handling irregular, memory-intensive dataflow patterns.

\section{Target Workload: Parallel-Sequential Patterns}
\label{sec:target}

\begin{figure}[t!]
	\centering
	\includegraphics[width=1\columnwidth]{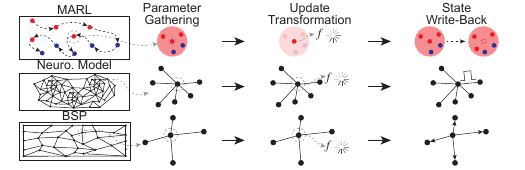}
	\caption{
		The three-stage processing of PSC tasks.
	}
	\label{fig:2_add}
\end{figure}

\begin{algorithm}[t!]
	\caption{Parallel-Sequential Computing Tasks}
	\renewcommand{\algorithmicrequire}{\textbf{Input:}}
	\renewcommand{\algorithmicensure}{\textbf{Output:}}
	\algdef{SE}[PARFOR]{ParFor}{EndParFor}[1]{\textbf{parfor} #1 \textbf{do}}{\textbf{end} \textbf{for}}
	\label{alg:1}
	\begin{algorithmic}
		\Require Entities $\vec{\mathbf{x}^{n}}(t)$, neighborhood $\mathcal{N}$, entity number $m$, persistent structural parameter $\mathbf{p}$
		\Ensure Final entity state $\vec{\mathbf{x}^{n}}(t_{end})$, total state number $t_{end}$.
		\For{each $t$}
		\For{each $i\in \left[0, m\right]$} \Comment{\textit{Parameter Gathering}}
		\State $\mathcal{N}_i = \text{FindNeighbor}(\vec{\mathbf{x}^{n}_{i}}(t))$
		\State ${\mathcal{D}}_{i} = \text{GBAccess}(\mathcal{N}_{i})$
		\EndFor
		\ParFor{each $i\in \left[0, m\right]$} \Comment{\textit{Update Transformation}}
		\State $\vec{\mathbf{y}^{n}_{i}}(t)={f}_{i}\left(\vec{\mathbf{x}^{n}_{i}}(t), \mathcal{D}_{i},\mathbf{p}\right)$
		\EndParFor
		\For{each $i\in \left[0, m\right]$} \Comment{\textit{State Write-Back}}
		\State $\vec{\mathbf{x}^{n}_{i}}(t + \Delta t) = \vec{\mathbf{y}^{n}_{i}}(t)$
		\Comment{\textcolor{gray}{Multicore Cache Coherence Sync.}}
		\EndFor
		\EndFor
		\Return$t_{end}$
	\end{algorithmic}
\end{algorithm}

\paragraph{PSC Formulation}
As depicted in Figure~\ref{fig:2_add} and Algorithm~\ref{alg:1}, PSC follows parallel-sequential patterns, which can be characterized by an evolving state space composed of entity-specific data.
To formalize the hardware requirements of these tasks, assume the global system state at any temporal index (``time'' or ``steps'') $t$ as $S_t$.
Each entity $i$ within this state is represented by a multi-dimensional feature vector $\mathbf{x}^{n}_{i}(t)\in \mathbb{R}^{n}$.
The persistent structural constraints of the task, such as graph topology, environmental boundaries, or synaptic weights, are encapsulated in the parameter set $\mathbf{p}$.
The state transition logic is governed by an update function $f_i$, which determines the feature vector of entity $i$ at state $t + \Delta t$ based on features of its neighborhood $\mathcal{N}_{i}$ and itself:
\begin{equation}
    \vec{\mathbf{x}^{n}_{i}}(t + \Delta t) = {f}_{i}\left(\vec{\mathbf{x}^{n}_{i}}(t), \{\vec{{\mathbf{x}_{j}^n}}(t)\}_{j\in \mathcal{N}_{i}},\mathbf{p}\right), i=1, 2,\dots,m
    \label{eq:1}
\end{equation}
where $\Delta t$ represents the step size between different states; $\mathcal{N}_{i}$ denotes the set of indices for entities influencing the update function of the $i^{\textrm{th}}$ entity; $m$ is the total entity population count.
From an architectural perspective, the primary overhead originates from the execution of $f$, which we decompose into three stages called \textit{Parameter Gathering}, \textit{Update Transformation}, and \textit{State Write-Back}.



\paragraph{Bottleneck Analysis}
%
We revisit Gustafson-Barsis's law~\cite{gustafson1988reevaluating} through a memory-centric analysis. According to Gustafson's formulation, the theoretical speedup \( S \) achieved via parallel computing is
\begin{equation}
	S = s + (1-s)N = N - (N-1)s,
	\label{eq:gustafson_speedup}
\end{equation}
where \( N \) denotes the number of processors and \( s \) represents the sequential fraction of the total execution time.

Consider an ideal PSC task described with Equation~\ref{eq:1}, in which \( m \) independent entities are processed concurrently across \( N \) cores, each requiring compute latency \( T_\text{1-step}^\text{compute} \). The sequential overhead arises from synchronization operations, specifically, gathering and writing back results. It incurs a data-transfer latency
$
T_\text{data-transfer} = \frac{C_\text{Gather} + C_\text{Back}}{BW},
$
where \( C_\text{Gather} \) and \( C_\text{Back} \) denote the data sizes for synchronization among the \( m \) cores, and \( BW \) is the global memory bandwidth. Hence, the sequential fraction \( s \) in Equation~\eqref{eq:gustafson_speedup} can be approximated as
$
s = \frac{T_\text{data-transfer}}{T_\text{data-transfer} + T_\text{1-step}^\text{compute}}.
$

Substituting this into Equation~\eqref{eq:gustafson_speedup} yields the memory-aware speedup approximation:
\begin{equation}
\begin{aligned}
		S &= N - (N-1) \frac{1}{1 + \frac{T_\text{1-step}^\text{compute}}{T_\text{data-transfer}}} \\
		&= N - \frac{N-1}{1 + \frac{T_\text{1-step}^\text{compute} \cdot BW}{C_\text{Gather} + C_\text{Back}}}.
\end{aligned}
	\label{eq:gustafson_speedup_approx}
\end{equation}

When memory access becomes the dominant bottleneck, i.e., $ \frac{T_\text{1-step}^\text{compute}}{T_\text{data-transfer}} \ll 1 $ due to relatively low bandwidth, the overall speedup \( S \) collapses to ``1'', indicating no effective parallelism gain.

\paragraph{Challenges.}
Based on the preceding analysis, memory access readily becomes the dominant performance factor due to two interconnected architectural challenges.

\textbf{Challenge 1: Global Buffer Saturation}
As shown in Figure~\ref{fig:1}(b), memory access dominates the execution profile, highlighting the significant communication overhead from $C_{Gather}$ and $C_{Back}$. The absence of direct peer-to-peer (P2P) communication primitives between processing elements in conventional accelerators means that scaling $m$ forces multiple cores to contend for global memory resources. This contention transforms the centralized GB into a serial bottleneck. Furthermore, in compute-centric architectures, the runtime dependency of $\mathcal{N}_{i}$ on entity states incurs substantial over-fetch overhead. The hardware must fetch entire cache lines to access non-contiguous entity data, resulting in severe bandwidth underutilization.

Although the \textit{Update Transformation} phase can be parallelized for SIMD/SIMT acceleration~\cite{lindholm2008nvidia, flynn2009some}, its low arithmetic intensity, often comprising only a few linear layers or element-wise nonlinearities, prevents effective utilization of computational units. Consequently, processing elements frequently stall waiting for data from saturated memory operations, leading to the second challenge.

\textbf{Challenge 2: The Memory-Bound Trap}
Figure~\ref{fig:1}(c--d) illustrates that regardless of workload complexity, GPU compute utilization remains low in PSC tasks. This indicates the bottleneck is an architectural latency issue, not a lack of raw computational throughput. The inefficiency originates from the rigid coupling of data indexing and processing in conventional architectures, which enforces sequential execution for irregular memory accesses. As a result, most of the processing time $T_{1-step}^{compute}$ in Equation~\ref{eq:gustafson_speedup_approx} is spent stalled, awaiting data fetches from the centralized global memory.

%

\paragraph{Motivation}
Building on the challenges outlined above, we introduce two key solutions in \codename:

\textbf{Multi-Tiered Peer-to-Peer Interconnect} (Section~\ref{sec:arch}):
\uline{Our core idea is to parallelize the original sequential part $T_{data-transfer}$ with a novel programming model and reconfigurable hardware substrate.} \dj{\codename employs a multi-tiered peer-to-peer communication fabric which filters the data communication according to the transmission distance, thereby eliminating contention for higher-level transmission paths.} By removing this centralized bottleneck, \codename avoids memory-access serialization and supports near-linear scalability.

\textbf{Data-Driven Programming Model} (Section~\ref{sec:model}--\ref{sec:comm}):
\dj{The proposed data-driven model replaces sequential control flow with fine-grained, data-driven memory operations.} Communication between cores is realized through memory-mapped I/O (MMIO) FIFOs, decomposing inter-core data transfers into discrete asynchronous events across the fabric. \dj{Consequently, these operations are hidden from the pipelines of each core, improving data independence of these cores and thus sustaining high computational throughput.}



\section{\codename Architecture}\label{sec:arch}

\begin{figure*}[!t]
	\centering
	\includegraphics[width=1\textwidth]{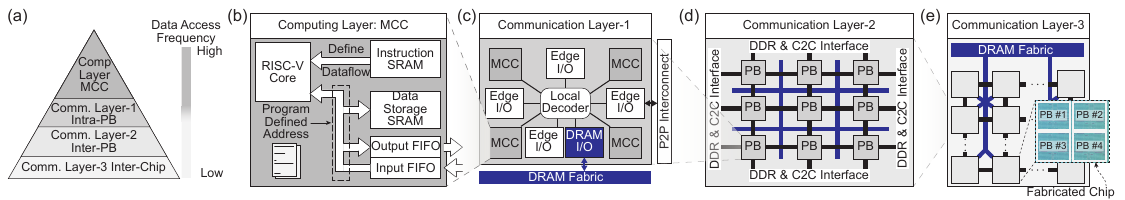}
	\caption{
		(a) \codename Communication ``Pyramid'';
        Architecture of
        (b) Computing Layer;
        (c,d,e) Communication Layer-1, -2, \& -3. Embedded figure in (e): \codename die photo for silicon verification.
	}
	\label{fig:2}
\end{figure*}

The \codename architecture employs a computing layer of Memory-Centric Cores (MCCs) interconnected by a multi-tiered NoC.
This design provides high-bandwidth data transmission while bypassing traditional multi-level cache structures to minimize multi-core cache coherence overhead.
The NoC fabric is organized based on spatial access frequencies to maximize data locality.
\dj{Compared with Cerebras~\cite{lie2023cerebras}, which implements data locality by holding everything on-chip, a three-tier implementation is illustrated in Figure~\ref{fig:2}(a), providing the basis for our architectural analysis and performance evaluation.}
The fabricated chip die photo of our implementation is embedded in Figure~\ref{fig:2}(e).

\subsection{Computing Layer: Memory-Centric Cores}
The fundamental component of \codename is the MCC, which augments on-chip memory with localized processing capabilities to minimize data movement overhead (Figure~\ref{fig:2}(b)).
To ensure deterministic pipeline control, each MCC integrates a RISC-V core as an embedded orchestrator, supported by dual SRAM banks dedicated to instruction and data storage, respectively.
Consequently, MCCs serve as the primary repositories for entity states $\mathbf{x}_i$, facilitating direct inter-core communication and bypassing the centralized GB and broader memory hierarchy.
Inter-core communication is managed via MMIO FIFOs with explicit programs.
This approach simplifies the programming abstraction while effectively masking data transmission latency within the processing pipeline.

\subsection{Communication Layer-1: Processing Block}\label{sec:PB}

In \codename, MCCs are grouped into Processing Blocks (PBs) to handle resource-intensive update functions $f$.
When a parallel-computed function (e.g., a full matrix multiplication) exceeds single-core capabilities, a PB facilitates collaborative execution through \codename Communication Layer-1.
This layer ensures data coherence between neighboring cores and manages the frequent data traffic resulting from multi-core collaboration.
Intra-block communication is supported by high-bandwidth FIFOs that enable near-instantaneous transfer of intermediate results.
Thus, during multi-core collaboration, entity state $\mathbf{x}_i$ remains resident in local memory, maximizing data locality.
Additionally, P2P interconnection between PBs and an independent DRAM Fabric is embedded in this layer to facilitate multi-block collaboration and efficient DRAM access (Figure~\ref{fig:2}(c)).

\subsection{Communication Layer-2 \& 3: 2D-Mesh}
The Communication Layer-2 and 3 of \codename consists of a 2D-mesh topology formed by P2P interconnection linking neighboring PBs (Figure~\ref{fig:2}(d--e)) and high-speed chip-to-chip (C2C) interface across multi-chip boundaries (e.g., using UCIe or PCIe protocol~\cite{mayhew2003pci, sharma2022universal}).
This distributed mesh serves as an active buffer, replacing multi-level caches to avoid redundant data movement caused by cache coherence across the memory hierarchy in multicore systems.
The mesh enables P2P communication to resolve neighborhood $\mathcal{N}_{i}$ memory queries and other essential inter-core data access without the serialization overhead of centralized memory controllers.
These links ensure system-wide data transparency, permitting the spatial distribution of complex functions $f$ (Algorithm~\ref{alg:1}) across multiple PBs or chips when necessary.
A dedicated DRAM Fabric handles off-chip traffic independently of the inter-core mesh.
This decoupling isolates the processing fabric from high-latency off-chip memory access and avoids DRAM data from being transmitted across the whole multi-tiered network to improve DRAM access efficiency.

\section{Data-Driven Programming Model}\label{sec:model}

\begin{figure}[t]
	\centering
	\includegraphics[width=1\columnwidth]{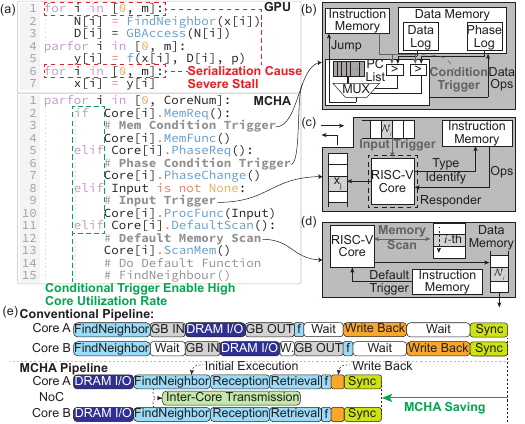}
	\caption{
		(a) Compute-centric and data-driven programs;
		(b) Dataflow for Conditional Trigger and Phase Conditional Trigger;
		(c) Dataflow for Input Trigger;
		(d) Dataflow for Default Memory Scan;
        (e) Pipeline Comparison of Conventional architecture and \codename.
	}
	\label{fig:3}
\end{figure}

Beyond conventional cache coherence management overhead, the performance of PSC workloads is limited by stalls arising from the tight coupling of data indexing and processing, which is a byproduct of the explicit data-fetch from memory to the central execution unit in the compute-centric model.
\textbf{We address this through a Data-Driven Programming Model that achieves latency hiding by breaking data access into a series of asynchronous events.}
By spreading these events across the distributed memory fabric, \codename allows data transmission to be overlapped with local processing, significantly reducing the impact of memory-related stalls.

Figure~\ref{fig:3} shows the programming model for \codename.
To enable data-driven processing, we categorize architectural operations of functions into four primary triggers:
\begin{enumerate}[leftmargin=*]
    \item \textbf{Memory Conditional Trigger}: Initiates functions when local SRAM variables meet logical predicates, enabling state-dependent data processing.
    \item \textbf{Phase Conditional Trigger}:
    Switches between triggering configurations at synchronization points to facilitate dynamic function deployment and smooth state transformations.
    \item \textbf{Input Trigger}: Activates functions upon message arrival in hardware FIFOs, decoupling execution from NoC latency and ensuring data-ready activation.
    \item \textbf{Default Memory Scan}: Performs an autonomous scan of local SRAM to trigger execution based on resident data in the absence of external FIFO inputs.
    This enables the MCC to initiate the computing loop without external signaling.
\end{enumerate}
The corresponding dataflows of these triggers within MCCs are presented in Figure~\ref{fig:3}(b--d).
A more detailed explanation and realization of our design is provided in Section~\ref{sec:flow}.

Furthermore, within this framework, to decouple data indexing from the processing pipeline, inter-core communication is decomposed into a three-stage asynchronous trigger sequence:
\begin{itemize}[leftmargin=*, label=$\diamond$]
    \item \textit{Initial Execution}, which triggers the generation and transmission of a data index, demonstrated by any trigger;
    \item \textit{Index Reception}, which initiates a local memory search and subsequent message return, demonstrated by the \textbf{Input Trigger};
    \item \textit{Data Retrieval}, where the arrival of the indexed data resumes the suspended computation, demonstrated by the \textbf{Input Trigger}.
\end{itemize}
By partitioning communication into these discrete asynchronous triggers, \codename effectively masks data transmission latency by overlapping it with the active processing pipeline.

Figure~\ref{fig:3}(e) presents the conceptual dataflow of programs in Figure~\ref{fig:3}(a).
In a conventional pipeline, the centralized GB obstructs the execution flow, leading to severe memory-bound performance degradation.
In contrast, by utilizing the \textbf{Default Memory Scan} trigger, MCCs initiate parallel off-chip data loading to maximize DRAM bandwidth utilization through hardware-level concurrency.
Computation of the updating function $f$ will be triggered by \textbf{Memory Conditional Trigger}.
Detecting the completion of the asynchronous data indexing process and stage transmission of the pipeline is handled through the \textbf{Phase Conditional Trigger}.
\dj{The implementation of the trigger-based data-driven programming model isolates the computing pipeline of each MCC, reducing $T_{data-transfer}$ in PSC workloads thanks to the elimination of stalls.}

\section{Communication Protocols}\label{sec:comm}

\dj{The design target of the programming model in Section~\ref{sec:model} is to enhance data independence of MCCs for isolation of the local pipeline from being obstructed by others.
To realize asynchronous communication for this purpose, we formalize the communication protocols of \codename with the help of MMIO FIFOs embedded in the MCCs instead of relying on a centralized memory controller like NVIDIA Hopper GPU~\cite{choquette2023nvidia}.
The direct mapping of the NoC interface into the local address space of each core eliminates the control-plane distinction between computation and communication, isolating the data access in each MCC from that of others.}
This design hides communication latency within the execution pipeline, which is particularly advantageous for PSC patterns.
Furthermore, the multi-tiered communication hierarchy mitigates bandwidth contention for long-distance transfers (e.g., inter-chip and off-chip data access) by isolating local traffic from global data movement.

\subsection{Core Message Management}\label{sec:core}

\dj{In \codename, each MCC functions as a self-contained execution environment where operations are triggered asynchronously by incoming messages or local memory states, consistent with the principles detailed in Section~\ref{sec:model}.}
Workloads are deployed to an MCC's local independent instruction memory with optimized execution flow from compilation.
As illustrated in Figure~\ref{fig:4}(a), communication FIFOs bypass complex network drivers and are instead mapped directly to the local address space.
Consequently, transmitting data to a neighboring MCC is compiled to a standard memory-store (\texttt{store}) operation.
When the RISC-V core writes to a designated output address, the data is enqueued into the output FIFO.
On this other side of receiving, data reception is handled via memory-load (\texttt{load}) operations, where the input FIFO functions as the core’s event queue for incoming messages awaiting processing.
\dj{The implementation of the MMIO FIFOs masks data transmission latency within the MCC execution pipeline, isolating the MCC local processing from the global NoC data transmission.}

\subsection{Transmission Message Definition}\label{sec:mess}

\dj{To shield the local MCC pipeline from unexpected stalls induced by the multi-tiered NoC latency, we isolate MCC execution from the global dataflow through a latency-hiding communication framework.
This framework stratifies data traffic based on transmission frequency and bandwidth availability, orchestrated by a streamlined 32-bit packet format.
By supporting two hardware-optimized message types that match the spatial locality patterns in \codename (Figure~\ref{fig:4}(b)), the design effectively minimizes routing overhead while ensuring high flexibility for intertwined PSC execution.}

Type~1 serves high-throughput, one-hop communication, corresponding to Communication Layer-1.
It constitutes the dominant traffic during the \textit{Gather} phase of Algorithm~\ref{alg:1}, where MCCs exchange entity states with adjacent PBs or neighbors in the 2D-mesh.
Each Type~1 message is a single self-contained 32-bit packet.
The first 3 bits serve as the Type ID and specify the hop direction (North, South, East, West, or Local).
The subsequent 5 bits in the same byte define the target (Cores or DRAM), leaving a 3-byte payload for user-defined data.
By restricting communication to a single hop, the hardware avoids complex routing headers and multi-cycle state tracking, enabling near-instantaneous data-residency updates.

Type~2 supports lower-frequency global communication using variable-length bursts, corresponding to Communication Layer-2 and 3.
Each burst comprises a single header followed by a specified number of data packets.
As with Type~1, the initial 3 bits identify the packet type.
A dedicated length field (bits 27--28 in Figure~\ref{fig:4}(b)) specifies the total number of subsequent packets in the burst.
The remainder of the header carries the routing information required to traverse the 2D-mesh and reach distant chip-level coordinates.
To maximize effective bandwidth, only this initial packet contains routing metadata; all following packets are devoted almost entirely to user-defined content, thereby enabling efficient transmission of weights, environmental parameters, or extended entity histories.

The entire transmission process is software-configurable via the RISC-V ISA.
This dual-tier protocol ensures that local neighborhood exchanges are resolved with minimal latency, while Type~2 messaging provides the necessary transparency for global MCC interactions across a large-scale multi-chip fabric.

\subsection{Intra-Block Transmission}\label{sec:intra-block}

\begin{figure}[tb]
    \centering
    \includegraphics[width=1\columnwidth]{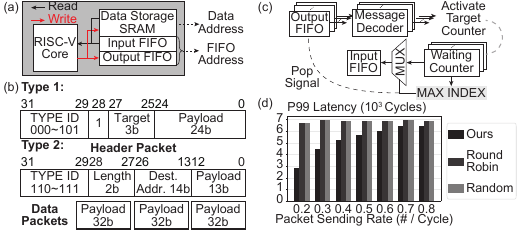}
    \caption{
        (a) MCC Message management method;
        (b) Message Packet Definition;
        (c) PB transmission strategy;
        (d) Ablation study comparing the P99 latency with other routing methods.
    }
    \label{fig:4}
\end{figure}

\dj{Multi-core collaboration within the PB triggers intense P2P data sharing, characterized by high-frequency, fine-grained message inquiries.
To support this behavior, we deploy the optimized Type~1 transmission format in Section~\ref{sec:mess}.
To realize stall-free coordination, PBs incorporate a high-bandwidth bidirectional interconnect linking the MCCs, as shown in Figure~\ref{fig:4}(c), enabling intra-block communication without traversing the global 2D-mesh network.
This localized routing effectively unburdens the global mesh and prevents cascaded core stalls.
}
In place of the power-intensive crossbars in general-purpose NoCs, the design employs a FIFO-based queuing and arbitration scheme optimized for PSC workloads.
At the output stage of each MCC or PB boundary, a hardware message decoder (Figure~\ref{fig:4}(c)) parses the Type ID and routing metadata of outgoing packets (Section~\ref{sec:mess}).
This decoder translates high-level message definitions into physical control signals that drive internal multiplexers: for Type~1 messages, it identifies the single-hop destination port; for Type~2 messages, it locks the path for the duration of a packet burst, ensuring contiguous transmission without interleaving jitter.

Resource contention arises when multiple MCCs attempt to transmit to the same destination concurrently.
Following the first- come-first-served principle inherent to FIFO-based designs~\cite{dally2004principles}, each input interface is equipped with a dedicated counter.
If an incoming message request is not accepted by the target FIFO, its counter increments whenever the target FIFO attempts to fetch messages.
During each arbitration cycle, the selector evaluates all active requests and grants priority to the message with the largest counter value—indicating the longest waiting time—pushing it into the input FIFO.
Once transmission begins, the corresponding counter is reset.
This policy serves as a critical anti-starvation mechanism, preventing indefinite delays in state updates.

An ablation study comparing the proposed arbitration policy with conventional Round-Robin and random-selection schemes is presented in Figure~\ref{fig:4}(d).
The simulation configures 8 input-output FIFO pairs, each with 1,000 packets destined for randomly chosen targets.
For the evaluated PSC tasks like MARL, long latency tails directly impact sequential processing speed and MCC stalling time; thus, we compare the latency at the 99th percentile (P99 latency) across the three policies.
With the simulated packet sending rate (x-axis) increase, the P99 latency (y-axis) increases to saturation.
This demonstrates that the first-come-first-served policy reduces the tail latency by 6.26\% to 41.8\% depending on packet generation rate, confirming the effectiveness of the selected design.

\subsection{Global Neighboring Connection}

\begin{figure}[tb]
	\centering
	\includegraphics[width=1\columnwidth]{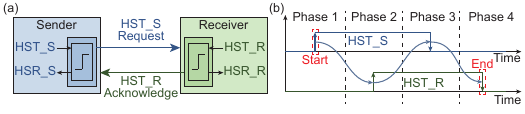}
	\caption{
		(a) Port definition for 1 internal connector pair
		(b) Timing diagram of the connectors in (a)
	}
	\label{fig:5}
\end{figure}

\dj{Following the same design philosophy of Section~\ref{sec:mess} and~\ref{sec:intra-block}, we decouple timing constraints across the \codename with an asynchronous network between neighboring PBs to achieve independent, self-governing execution.}
This allows each PB to operate within its own independent clock domain, reducing global clock distribution complexity and mitigating peak power surges.
As shown in Fig.~\ref{fig:5}, inter-block data transfers (e.g., Block~A to Block~B) are governed by a robust 4-phase asynchronous handshake protocol:
\begin{enumerate}[leftmargin=*, label=Phase \arabic*.]
	\item \textit{Request}: The protocol begins with Block~A driving its handshake sending port (HST\_S) high to signal a request.
	\item \textit{Acknowledgement}: Upon detecting this request, Block~B responds by driving its acknowledgement port (HST\_R /HSR\_S) high.
	\item \textit{Clear Request}: Block~A then clears its request signal and reset HST\_S to ``0''.
	\item \textit{Clear Acknowledgement}: After which Block~B clears its acknowledgement, completing the cycle and freeing the channel for the next transfer.
\end{enumerate}
This lightweight handshake acts as a hardware-level flow control, ensuring data residency and integrity across the mesh even when neighboring blocks operate at different frequencies or states.

When scaling to multi-chip systems, \codename extends its 2D-mesh communication protocol across chip boundaries using high-bandwidth interconnects such as PCIe~\cite{mayhew2003pci} or UCIe~\cite{sharma2022universal}.
These system-level links maintain information transparency, enabling a distant core on a remote chip to be accessed via the same asynchronous handshake strategy.
Consequently, \codename provides a modular and scalable fabric capable of supporting PSC across multiple chips.

\section{System Deployment and Mapping Strategy}\label{sec:deploy}
\dj{This section elaborates how to maximize the communication efficiency of the multi-tiered NoC and the parallelism of MCCs through the formulation of algorithms along with their deployment strategies on the \codename platform.}
Figure~\ref{fig:6}(a--b) illustrates the programming and deployment workflow of \codename with a representative program example, matrix processing $\mathbf{A}\cdot(\mathtt{prod}(\mathbf{B}))-\mathbf{C}$, where \texttt{prod}() computes the product of all numerical elements in a matrix.

\subsection{Collective Usage of MCC Buffers}\label{sec:mem}

\dj{
Since \codename enables high independence to the MCCs, the management of the on-chip memory and workload varies from the conventional systems.
Instead of allocating specific tasks to different cores, \codename maps computational entities to cores based on the memory footprint of their state data (\( {\mathbf{x}}_{i}^{n} \) in Algorithm~\ref{alg:1}).
Only cores that control this data can modify it; others can either read the data or send write requests to these cores.
Compared with threads in a GPU that are assigned to specific operations each time (Figure~\ref{fig:6}(c--d)), cores in \codename perform different operations based on memory types to parallelize the irregular data access, namely $T_{data-transfer}$ in PSC workloads.
}
As presented in Figure~\ref{fig:6}(a), Core 0 and 2 are responsible for loading the data of matrices \textbf{A} and \textbf{C}, acting as the entity buffer.
When the state data size surpasses the capacity of on-chip SRAM, only the necessary indices for accessing off-chip DRAM are maintained in SRAM, as Core 0 indexing data of vector \textbf{A} in Phase 1.
By organizing these index structures in alignment with the underlying hardware topology (e.g., arrange the elements of \textbf{A} and \textbf{C} alternately in the DRAM), we convert the serialized and irregular memory access characteristics of traditional paradigms into predictable, parallelized local transactions.
Consequently, the most frequently accessed variables remain physically co-located with the corresponding processing logic, thereby eliminating the ``memory wall'' at the architectural level.

\subsubsection{Locality-Aware Entity Clustering}

\dj{
To filter local data transfers through the multi-tiered NoC, thereby minimizing communication overhead in the 2D-mesh, the mapping process prioritizes computational locality by assigning neighboring entities to the same MCC or adjacent cores within a PB based on their interaction frequency.
For PSC workloads, we can use clustering~\cite{ikotun2023k} or graph-partitioning ~\cite{ccatalyurek2023more} algorithms to group entities with overlapping neighbor sets ($\mathcal{N}_{i}$ in Algorithm~\ref{alg:1}), thus maximizing data reuse within local SRAM banks.
}
In the provided example, this process is represented as placing the three cores within the same PB.
As the neighbor sets $\mathcal{N}_{i}$ evolve during task execution, e.g., due to agent movement necessitating inter-core communication, \codename supports periodic entity migration, with frequency and strategy tailored to the task.
In complex environments such as Multi-Agent Particle Environments (MPE)~\cite{mordatch2017emergence, peng2021facmac} or StarCraft maps~\cite{flair2024jaxmarl}, the MARL simulation space is initially partitioned into blocks (Figure~\ref{fig:6}(e)).
Throughout sequential steps in PSC patterns, these block boundaries are dynamically adjusted based on local agent density, rebalancing computational load and ensuring consistent architecture-wide performance.

\subsubsection{Systolic Mapping for Complicated Update Functions}

\begin{figure}[tb]
\centering
\includegraphics[width=1\columnwidth]{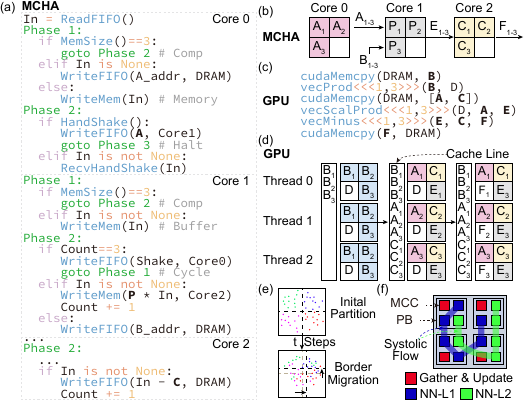}
\caption{
    (a) \dj{A program for} \codename;
    (b) Conceptual dataflow of (a);
    (c) Program using CUDA;
    (d) Conceptual dataflow of (c);
    (e) In-process update enabled by 2D-mesh;
    (f) On-chip systolic function demonstration;
}
\label{fig:6}
\end{figure}

\dj{When processing computationally intensive update functions $f$, such as deep neural network inference, $T_{data-transfer}$ is embedded in the data sharing between collaborating MCCs.
Workloads requiring frequent communication, for example, the lookup-table of non-linear functions and corresponding input generatos, are implemented within the same blocks.
Others are spread across the system according to the dataflow structure, formulating a systolic array.
}
For example, in Figure~\ref{fig:6}(a--b), Core 1 acts as the buffer of intermediate results for computing $\textbf{A}\cdot(\texttt{prod}(\textbf{B}))$.
Applying an input-stationary systolic mapping~\cite{lee2024resa}, these cores first store the input data (\textbf{A} in this example) and then, triggered by the completion of input transmission, network weights (\textbf{B} in this example) are streamed through these MCCs while previously prepared input data remain stationary in local SRAM.
By sequencing weights according to input channels, the architecture can skip null-value operations and support sparsity-aware computation.
In detail, \codename MCCs demonstrate this process through 2 continuous phases, one for temporal buffering (Phase 1 of Core 1), the other for data indexing and computing (Phase 2 of Core 1).
The completion of the input transmission acts as the \textbf{Phase Change Trigger} of Phase 1, changing the core to the computing phase, where data indexing of \textbf{B} is default.
The core then performs operations according to pre-defined instructions (\texttt{prod} in this example, multiply-and-accumulate for neural networks), triggered by the arrival of indexed data.
Figure~\ref{fig:6}(f) illustrates the systolic mapping of a 2-layer Neural Network (NN) in \codename.
The blue cores (labeled as NN-L1) store the data generated in the \textit{Parameter Gather} process, namely the input data of the first NN layer, acting as intermediate buffers (a similar role as Core 1 in Figure~\ref{fig:6}(b)).
This spatial unrolling prevents computationally dense transformation phases from becoming bottlenecks, thereby maintaining the efficiency of the data-driven computing flow.

\subsection{Data-Driven Programming Model Implementation by RISC-V ISA}
\dj{The reconfigurability of \codename is demonstrated through the programmable triggering logic based on standard RISC-V instruction set (RV32I as an example).
The ISA execution of each MCC is governed by the state of the FIFOs and local SRAM conditions through 4 different triggers in the programming model at each stage.
}

\subsubsection{MMIO-Based FIFO Interaction}
\dj{By leveraging the MMIO FIFOs described in Section~\ref{sec:core}, sending and receiving data in MCCs reduces to standard RISC-V \texttt{LW} (Load Word) and \texttt{SW} (Store Word) instructions. This decouples communication from computation, allowing each MCC's processing pipeline to remain uninterrupted and transmission latency to be effectively hidden.}
(a) When a compiled \texttt{SW} instruction targets the output FIFO address (\textcolor[HTML]{609cca}{\texttt{WriteFIFO}}), the hardware pushes the 32-bit payload into the transmission queue.
To prevent data loss, if the FIFO is full, the hardware interface stalls the pipeline or maintains the value in the register file until a slot becomes available, similar to the behavior of memory handshake in conventional \texttt{SW} operations.
Crucially, the ISAs ensure atomic transmission; once a multi-packet message (Type 2) begins, the hardware locks the port to ensure the packets are transmitted contiguously without interleaving jitter.
(b) An \texttt{LW} instruction targeting the input FIFO (\textcolor[HTML]{609cca}{\texttt{ReadFIFO}}) pops the front-most 32-bit packet.
To enable high-speed triggering, the hardware utilizes a zero-signature mechanism: if the FIFO is empty, the \texttt{LW} returns a 32-bit ``0'', which triggers the \textbf{Default Memory Scan} if it exists.
Conversely, if a valid packet is present, the protocol guarantees that at least one bit is non-zero (bit 28 for Type 1 or bit 31 for Type 2, as shown in Figure~\ref{fig:4}(b)).

\subsubsection{Control Flow Instruction Mapping}\label{sec:flow}
As presented in Figure~\ref{fig:6}(a), each time MCC is idle, it will read the address corresponding to the input FIFO if \textbf{Input Trigger} is not empty and check all the other triggers.
Thus, the basic structure of the program at each phase is a loop, which repeats the data detection process until jumping into other phases.
The 4 triggering conditions are accomplished by RISC-V branching instructions, as demonstrated by the if statement in Figure~\ref{fig:6}(a), allowing the core to react to the memory fabric in real-time.
(a) \textbf{Memory \& Phase Conditional Triggers}: These two kinds of triggers rely on reconfigurable data statistics stored at pre-programmed addresses within the local SRAM.
By periodically reading these status variables, the RISC-V branch logic determines if the core should handle special operations or switch functional phases.
(b) \textbf{Input Trigger}: By executing an \texttt{LW} from the input FIFO followed by a conditional branch, such as \texttt{BNE}, the core can immediately detect incoming data.
If the result is non-zero, the Program Counter (PC) jumps to the corresponding processing function.
(c) \textbf{Default Memory Scan}: If an \texttt{LW} from the input FIFO returns zero (empty FIFO), the core can default to a memory scan.
This is managed via a scan pointer maintained in the register file, which incrementally traverses the local SRAM to check for a response.
Ultimately, every functional trigger in the \codename model is represented as a PC jump to a specialized firmware routine.
\dj{By mapping the programming model components into existing RISC-V instructions, \codename introduces minimal efforts for ISA construction.}

\begin{table}
\caption{\dj{Design Parameters for Large-Scale Simulation}}
\footnotesize
\begin{tabulary}{\columnwidth}{L|L|C}
\noalign{\hrule height 1pt}
 Part                              & Parameter                   & Value    \\
\hline
\multirow{4}{*}{MCC}           & Memory Size (KB)                  & 1              \\
                               & Pipeline Parallelism         & 1                \\
                               & FIFO Buffer                  & 8                \\
                               & Processing Throughput (GFLOPS/Core)            & \dj{1.6}  \\
\hline
\multirow{4}{*}{PB}            & Edge Connector               & 4             \\
                               & DRAM Connector               & 1             \\
                               & Number of Cores              & 4             \\
                               & Connector Bandwidth (GB/s)          & 1024     \\
\hline
\multirow{5}{*}{\codename}     & External Connector           & 4             \\
                               & DRAM Access Parallelism      & 16            \\
                               & Number of Blocks             & 16            \\
                               & External Connector Bandwidth (GB/s) & 64        \\
                               & DRAM Access Bandwidth (GB/s)        & 409.6    \\
\noalign{\hrule height 1pt}
\end{tabulary}
\label{tab:1}
\end{table}

\subsection{Dataflow-Defined Workload Assignment}
\dj{To maximize \codename throughput, the assignment of the workloads should involve identifying essential reactive elements and mathematically modeling throughput requirements to determine the optimal pipeline depth for the target computing tasks.}

\subsubsection{Functional Requirements for Data-Driven Programs}
\dj{
Similar to object-oriented programming, the \codename requires programmers to break the target algorithm into multiple statically defined phases.
The 4 types of triggers of a specific phase act as the member functions of an object.}
The programmer should define \textbf{Phase Conditional Trigger}, which serves as the termination condition for a computational loop, analogous to loop-exit logic in conventional paradigms.
Furthermore, each phase must include at least one definition of an \textbf{Input Trigger} or a \textbf{Default Memory Scan} to dictate the specific operations dominating that phase (\textcolor[HTML]{609cca}{\texttt{WriteFIFO}} and \textcolor[HTML]{609cca}{\texttt{WriteMem}} in Figure~\ref{fig:6}(a) for Core 0).
If neither is defined, the MCC becomes a ``zombie'' core, because it lacks both external stimulus responses and autonomous self-activation.
\dj{The transmission from the conventional programs to \codename ones can be accomplished with the help of Large Language Models (LLMs).
We provide an example that transforms a neural network to \codename programs in C++ with the help of Deepseek-R1~\cite{guo2025deepseek} in thinking mode in our github (\url{https://github.com/carabdis/MCHA/LLM.md}).}

The program must explicitly include handshake signal generation to manage the flow of data across the pipeline, ensuring that residency is maintained without causing mesh-wide stalls.
\dj{As presented in Figure~\ref{fig:6}(a), Phase 2 of Core 0 and 1, the explicit handshake acts as the data barrier for the serialized part in PSC workloads.}

\subsubsection{Analytical Deployment and Throughput Matching}\label{sec:match}
\dj{After decoupling the PSC workloads into serialized phases, the deployment of the algorithm can be defined as a flow matching problem. We provide a 3-step analytical process to deploy a program onto \codename fabric to balance the workload across the system.}

\textbf{Step-1 Throughput Extraction}: The processing characteristics of MCC can be described through two critical metrics, consumption rate $R_{i}^{pop}$ and production rate $R_{i}^{push}$.
$R^{pop}$ is defined as the average cycles required for a core to pop and parse a 32-bit input packet from the input FIFO.
$R^{push}$ is defined as the average cycles required for the core to compute the assigned workload and push the resulting packet to the output FIFO.
\dj{Thus, the goal is to match the ingress and egress rates of each stage in the pipeline.}
For example, neglecting the DRAM access latency, Core 0 in Figure~\ref{fig:6}(a) generates 1 output packet per cycle in Phase 2, namely $R_{0}^{push}=1$.
Core 1 consumes the 1 input packet from Core 0 every 3 cycles to compute the intermediate results of $\textbf{A}\cdot(\texttt{prod}(\textbf{B}))$, thus $R_{1}^{pop}=1/3$.

\textbf{Step-2 Ideal Pipeline Derivation}: \dj{Under the assumption of infinite hardware resources, the ideal parallelized core count for each phase ($N_i^{ideal}$) required to achieve a non-blocking pipeline without hardware waste is derived from the ratio of these rates.}
To maintain a steady state where FIFOs neither overflow nor starve, the flow balance between serialized phases is defined as:
\begin{equation}
    N_{i}^{ideal} * R_{i}^{push} = N_{i - 1}^{ideal} * R_{i}^{pop}
\end{equation}
In our example, since $R_{0}^{push}=1$ and $R_{1}^{pop}=1/3$, the ideal ratio $N_{0}^{ideal}:N_{1}^{ideal}$ is $1:3$.
This ratio determines the quantitative relationship between MCCs operating in neighboring phases, minimizing unnecessary data movement of intermediate results.

\textbf{Step-3 Realistic Constraint Scaling}:
Given a fixed physical scale, the deployment logic determines the maximum number of concurrently active phases \( M_{\text{para}}^{\max} \). Due to memory structural hazards—where buffering and computation cannot occupy the same SRAM bank simultaneously—parallel phases must be interleaved with idle or buffering phases or employ twice the memory capacity via ping-pong buffering.
The algorithm computes \( M_{\text{para}}^{\max} \geq 1 \) based on the minimum buffer cores required per phase. In the former example, if phases require \(\{1, 2, 1\}\) memory cores for \( M_{\text{para}} = 2 \), but only 3 cores are available, then \( M_{\text{para}}^{\max} = 1 \). Starting from \( M_{\text{para}}^{\max} \), the system iteratively reduces the number of active phases until memory capacity constraints are satisfied. This scaling ensures robustness across implementations, from single-chip edge devices to large-scale multi-chip fabrics.

\section{Evaluation}
\subsection{Experimental Setup}
To evaluate the performance and efficiency of large-scale \codename architecture, we developed a cycle-accurate simulator.
This simulator is directly extracted from RTL Verilog HDL implementation, including MCCs, PBs, on-chip 2D-mesh design, and the chip-wise connectors.
The chip-wise connectors demonstrated in the Verilog code are SPI connectors as a research prototype, which are replaced by parameters of commercial PCIe 4.0 in the simulator.
\dj{Other hardware parameters are modeled based on logic synthesis results using TSMC 28nm PDK at 200MHz}.
\dj{Detailed parameters of one \codename chip using in the simulator are in Table~\ref{tab:1}}.
The corresponding area and power overhead of our implementation are given in Table~\ref{tab:2}.
\begin{table}
\caption{The power and area breakdown of \codename}
\footnotesize
\begin{tabulary}{\columnwidth}{L|L|C|C|C|C}
\noalign{\hrule height 1pt}
 Part       & Component
                        &Area (mm$^2$)
                        & Area (\%)
                        & Power (mW)
                        & Power (\%) \\
\hline
\multirow{5}{*}{MCC}    & SRAM          & 0.0303    & 82.78     & 0.4968     & 41.40      \\
                        & FIFO          & 0.0006    & 1.64      & 0.0706    & 5.88       \\
                        & RISC-V Core   & 0.0044    & 12.02     & 0.4913    & 40.94      \\
                        & Control       & 0.0013    & 3.55      & 0.1412    & 11.78      \\
                        & Total (MCC)   & 0.0366    & -         & 1.2000    & - \\
\hline
\multirow{3}{*}{PB}     & MCC           & 0.1464    & 94.57     & 4.8000    & 76.78      \\
                        & Control       & 0.0084    & 5.42      & 1.4514    & 23.21      \\
                        & Total (PB)    & 0.1548    & -         & 6.2514    & -  \\
\hline
\multirow{4}{*}{\codename}
                        & PB             & 2.4768    & 84.80     & 100.0224  & 86.70      \\
                        & Ext. Connector & 0.0560    & 1.92      & 1.9352    & 1.68       \\
                        & Int. Connector & 0.3878    & 13.28     & 13.4018   & 11.62      \\
                        & Total          & 2.9206    & -         & 115.3594  & - \\
\noalign{\hrule height 1pt}
\end{tabulary}
\label{tab:2}
\end{table}

To verify \codename, we compare the performance of the \codename system composed of 4 (\codename-4) and 32 (\codename-32) chips to various kinds of DSAs and generalized GPUs.
\textbf{NVIDIA A100 GPU} is used as the baseline for massive-scale parallel workloads with flexibility.
\dj{\textbf{PEARL}~\cite{li2025pearl}, \textbf{ActiveN}~\cite{liu2024activen}, \textbf{MC$^2$A}~\cite{zhao2025mc}, \textbf{Dalorex}~\cite{orenes2023dalorex}, and \textbf{PolyGraph}~\cite{dadu2021polygraph} are chosen as the specialized accelerators for different benchmarks.}
The evaluation covers three distinct domains, highlighting \codename's ability to handle various types of programs:
\begin{enumerate}[leftmargin=*,label=(\arabic*) ]
\item \textbf{MARL}: a comprehensive suite including \textit{MPE Simple Spread}~\cite{mordatch2017emergence}, \textit{StarCraft II (SMAX)}~\cite{flair2024jaxmarl}, \textit{Switch Riddle}~\cite{foerster2016learning}, and \textit{STORM}~\cite{lu2022model, foerster2018learning}.
Performance is compared against \textbf{JAX-MARL}~\cite{flair2024jaxmarl} running on \textbf{A100} or \textbf{RTX 3090} and \textbf{PEARL}~\cite{li2025pearl}.
For each environment, we measure the average processing speed of the baselines computing 1,000 timesteps.
The number of agents existing in these environments is fixed as 1,000, to manifest the efficiency in large-scale MARL computation.
To ensure a fair comparison, the \codename implementation utilizes the same hyperparameters as the JAX-MARL demonstrations, including neighborhood size $\mathcal{N}_{i}$, action space, and update functions$f_{i}$.

\item \textbf{Large-Scale Neuromorphic Computing}: The benchmark is a 2\% MVC (Motor Variable Control) nervous system model~\cite{schmidt2018multi}, which follows the setup established by \textbf{ActiveN}~\cite{liu2024activen}, testing the architecture's capability to handle sparse, high-frequency spikes and deterministic timing.

\item \textbf{Typical BSP for Graph Processing}: \dj{We evaluate \codename’s performance on regular data access patterns using PageRank~\cite{page1999pagerank}, BFS~\cite{murphy2010introducing}, and Markov Random Field (MRF) image segmentation~\cite{zhang2021statistical}, compared with existing DSAs (\textbf{MC$^2$A}~\cite{zhao2025mc}, \textbf{PolyGraph}~\cite{dadu2021polygraph}, and \textbf{Dalorex}~\cite{orenes2023dalorex}) and GPUs~\cite{zhao2025mc, zhao2024aia, ko20203mm, dadu2021polygraph}.}
\dj{The evaluation uses the \texttt{orkut} graph~\cite{mislove2007measurement} when comparing PageRank and BFS against PolyGraph~\cite{dadu2021polygraph}, and the \texttt{RMAT-26} graph~\cite{leskovec2010kronecker} when comparing against Dalorex~\cite{orenes2023dalorex}. MRF image segmentation input data follows~\cite{zhang2021statistical}.
}
\end{enumerate}

\dj{
For the above workloads, \codename matches the algorithmic setup of every corresponding baseline~\cite{flair2024jaxmarl, liu2024activen, dadu2021polygraph, orenes2023dalorex, zhao2025mc} in numerical precision, task size, step definition, and convergence condition.
}
\subsection{End-to-End Computing Throughput}

\subsubsection{Multi-Agent Reinforcement Learning (MARL)}

Figure~\ref{fig:7}(a) compares the absolute execution time (y-axis) of different hardware (x-axis), including \codename (M1, M2), GPUs (G1--G2), and FPGA (F1).
\dj{The presented throughput time is measured end-to-end including data off-loading and data processing.}
\codename-4 delivers 153.06$\times$ to 2456.96$\times$ speedups versus GPU-accelerated frameworks.
\codename-32 further achieves 1.22$\sim$4.25$\times$ speedup over \codename-4.
This performance leap stems from the elimination of redundant global memory copies due to frequent cache coherence and overfetch in GPUs.
Furthermore, the high data locality achieved in \codename in turn removes most of the irregular data access to DRAM, thus improving the overall performance.

Compared to PEARL (F1), \codename-4 achieves a speedup of 2.28$\times$ to 4.78$\times$.
This performance gain is primarily attributed to the data-driven adaptivity inherent to the \codename architecture, which masks data transmission latency across the multi-tiered NoC.
Furthermore, \codename-4 eliminates the significant communication overhead typically associated with data transfers between external GPUs and the FPGA fabric, streamlining the execution of PSC workloads.

\subsubsection{Large-Scale Neuromorphic Computing}
\dj{Figure~\ref{fig:7}(c)} shows the comparison results benchmarking \codename (M1, M2), neuromorphic accelerators (AN, NE, Loihi)~\cite{liu2024activen, davies2018loihi, lee2021neuroengine} and GPUs (G1, G2, G4).
\codename-4 shows 5.36$\times$ improvement over the neuromorphic-specific accelerators~\cite{liu2024activen, davies2018loihi}, completing a single task in 0.7842s, while the \codename-32 presents 39.996$\times$ speedup.
When compared with ActiveN, \codename-32 performs 1.175$\times$ speedup with 9.36\% power consumption, presenting the high efficiency of \codename originating from the data-driven nature.
Furthermore, the near-linear performance improvement between \codename-4 and \codename-32 originates from the high spatial locality achieved by the communication strategy (more details are investigated in Section~\ref{sec:scale}).

\subsubsection{Typical BSP for Graph Processing}
\begin{figure}[t!]
    \centering
    \includegraphics[width=1\columnwidth]{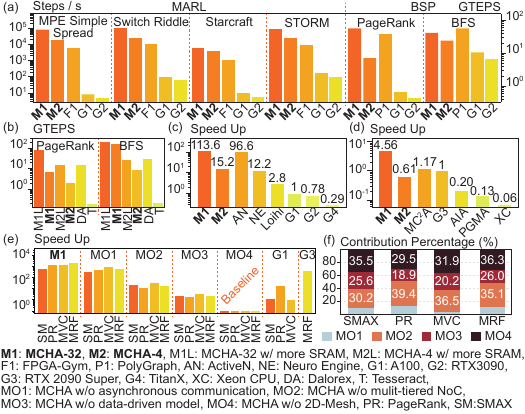}
    \caption{
        The speedup compared with the existing
        (a) MARL and BSP;
        \dj{(b) more BSP;}
        (c) MVC;
        (d) BSP-MRF accelerators.
        \dj{(e) The ablation study of \codename-4.
        (f) The detailed contribution of different characteristics of \codename.}
    }
    \label{fig:7}
\end{figure}

\dj{Figure~\ref{fig:7}(a), (b)\&(d)} presents the results.
\codename presents comparable results versus various DSAs for these applications.
Compared with general-purpose GPUs for BSP tasks, \codename-32 achieves 86.69$\times$, 3.93$\times$, and 4.56$\times$ speedup, respectively, while \codename-4 also presents comparable performance.
\dj{We further compare \codename with Dalorex\cite{orenes2023dalorex}, whose 256 cores match the compute-unit count of \codename-4 (``M2'' in Figure\ref{fig:7}). M1L and M2L denote M1 (\codename-32) and M2 (\codename-4) with on-chip memory scaled to 4.2,MB/core to match Dalorex's configuration, respectively. In their original setup (M1 and M2), \codename shows little advantage over Dalorex; however, M1L achieves a 5.66$\sim$7.28$\times$ throughput improvement, and M2L reaches throughput comparable to Dalorex. This gap is explained by DRAM access overhead: Dalorex's model excludes DRAM latency, whereas \codename's includes it. On the RMAT-26~\cite{leskovec2010kronecker} dataset, DRAM accesses consume 81.42\% of M1's execution time under the original 1KB/core configuration — an overhead largely eliminated by the enlarged on-chip memory in M1L.}

\dj{The lower speedup of BSP implementations relative to MARL and MVC stems from differences in arithmetic intensity across the workloads.}
The most informative metric is the bandwidth utilization rate.
In the BSP-MRF task, the performance ratio between \codename-4 and the GPU (0.61) is close to the ratio of their respective DRAM bandwidths (0.67).
This correlation proves that \codename-4 maintains a near-optimal bandwidth utilization rate in regular data processing tasks, \dj{while also presenting the drawbacks of \codename on computing-intensive tasks because of the limited scale.}

\dj{Figure~\ref{fig:7}(e) and (f) present an ablation study of \codename-32, starting from a bus-based architecture and incrementally adding one feature at a time until reaching the full \codename design. Each feature's contribution is quantified by the latency reduction it yields over the bus-based baseline on a given workload. The results show that the bulk of the speedup over GPUs stems from the \textbf{multi-tiered NoC} and the \textbf{data-driven model} proposed by \codename, with the asynchronous communication strategy providing additional gains. This breakdown confirms the effectiveness of \codename's core tenets, i.e., data localization and data-driven programming.
}

\begin{table}
\caption{\dj{The memory footprint of \codename-4}}
\footnotesize
\begin{tabulary}{\columnwidth}{L|C|C|C|C|C}
\noalign{\hrule height 1pt}
\dj{Benchmark}          & \dj{Core Memory Ops (MB)}
                        & \dj{Intra-Block (MB)}
                        & \dj{Inter-Block (MB)}
                        & \dj{Inter-Chip (MB)}
                        & \dj{DRAM Access (MB)}\\
\hline
\dj{MPE Simple Spread}  & \dj{211.900}  & \dj{146.152}  & \dj{67.318}   & \dj{0.000}    & \dj{0.092}    \\
\hline
\dj{Switch Riddle}      & \dj{102.355}  & \dj{64.161}   & \dj{30.685}   & \dj{0.000}    & \dj{0.035}    \\
\hline
\dj{Starcraft}          & \dj{1044.362} & \dj{461.526}  & \dj{141.667}  & \dj{0.854}    & \dj{0.139}    \\
\hline
\dj{STORM}              & \dj{158.167}  & \dj{107.375}  & \dj{40.645}   & \dj{4.378}    & \dj{0.077}    \\
\noalign{\hrule height 1pt}
\end{tabulary}
\label{tab:3}
\end{table}

\subsection{Architectural Analysis}
To elucidate the fundamental reasons for the observed \codename performance gains, we explore the architectural design space and profile the benchmark.
We first examine the simulation results of the end-to-end throughput measurement and find the components of the processing time of \codename-4.
\dj{We also provide the memory footprint of \codename-4 on MARL workloads.}
Based on these analyses, we formulate a customized roofline model, establishing the operational upper bounds for \codename.
Utilizing this model, we examine the scalability characteristics of \codename to identify the structural features that enable near-linear performance growth across increasing core counts and the variation of speedup factor between \codename-4 and \codename-32 in different cases.
Further, we conduct a sensitivity analysis, quantifying how variations in multi-level interconnect bandwidths and RISC-V compute intensity, as detailed in Table~\ref{tab:1}, impact the end-to-end throughput of PSC applications.

\subsubsection{Computing Intensity Analysis}\label{sec:intense}
\begin{figure}[t!]
    \centering
    \includegraphics[width=1\columnwidth]{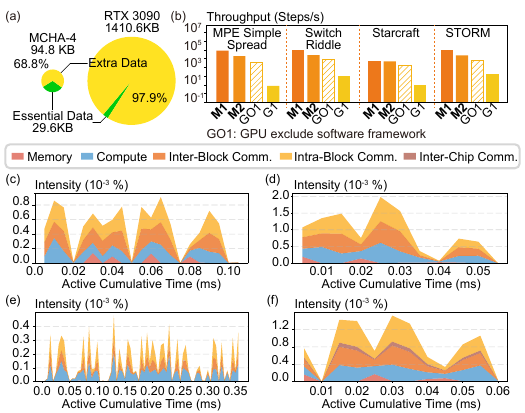}
    \caption{
        \dj{(a) The total data movement of \codename-4 and RTX3090.
        (b) The throughput of \codename-32, \codename-4 and A100 GPU with and without the software framework time costs.}
        The computing intensity of (c) MPE Simple Spread, (d) Switch Riddle, (e) Starcraft, and (f) STORM of \codename.
    }
    \label{fig:8}
\end{figure}

\dj{Figure~\ref{fig:8} shows the results of profiling runtime operational types for \codename-4. The x-axis is the execution time.}
The y-axis shows the operational intensity in percentage: ``compute'' for computation, ``memory'' for memory I/O access operations, ``Comm.'' for the three types of NoC operations.

A key insight from this breakdown is that these workloads are overwhelmingly communication-dominant.
\dj{Intra-block and inter-block data movements constitute 57.3\%$\sim$69.1\% of operational cycles, which underscores why conventional GPUs struggle with PSC tasks with the detailed memory footprints of \codename in Table~\ref{tab:3}}.
Lacking an efficient P2P fabric for inter-core data exchange, GPUs force high-frequency transmissions through the global memory hierarchy.
\dj{This can be illustrated by profiling the MPE Simple Spread benchmark on an RTX3090 with Nsight Systems. The essential data per agent amounts to 8 bytes, comprising the weight vector (hidden dimension 64) and the environment state. As shown in Figure\ref{fig:8}(a), the total off-chip memory traffic of the RTX~3090 running MPE Simple Spread with 1,000 agents over 1,000 steps is 14.88$\times$ larger than that of \codename-4, a disparity driven by cache-coherence traffic across L2 cache lines. The reason \codename-4's data access still exceeds the application's essential working set is the four-fold replication of the neural network, stored once per chip. Nevertheless, \codename-4's multi-tiered NoC preserves data locality and bypasses off-chip memory and the global buffer.
}
Compared to the GPU computing intensity in Figure~\ref{fig:1}(d) in Section~\ref{sec:intro}, where DRAM access accounts for 78.9\% of the total operations, DRAM traffic in \codename occupies only 2.26\%$\sim$5.44\%.
\uline{This significant reduction validates that \codename successfully transforms global memory bottlenecks into\\localized on-chip data traffic.}

\dj{To further demonstrate the efficacy of the \codename architecture, we compare the throughput of \codename-32, \codename-4, and an NVIDIA A100 on MARL workloads after factoring out GPU software framework overhead. Even after this normalization, \codename-32 still achieves 4.01$\sim$21.93$\times$ speedup. This overhead originates predominantly from XLA~\cite{sabne2020xla} just-in-time compilation, which, though intended to optimize memory-bound workloads, consumes 97.5\%$\sim$99.79\% of total GPU execution time. Such a stark imbalance underscores the advantage of the proposed data-driven programming model: it statically determines the instruction schedule from the base algorithm while sustaining high hardware utilization. The overlap of computation and data movement further validates the model's ability to hide transmission latency within the pipeline.
}

\begin{figure}[t!]
    \centering
    \includegraphics[width=1\columnwidth]{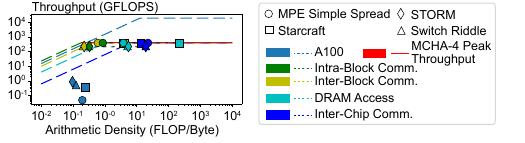}
    \caption{
        \dj{Roofline model analysis for \codename-4 and A100}.
    }
    \label{fig:9}
\end{figure}

\begin{figure}[t!]
    \centering
    \includegraphics[width=1\columnwidth]{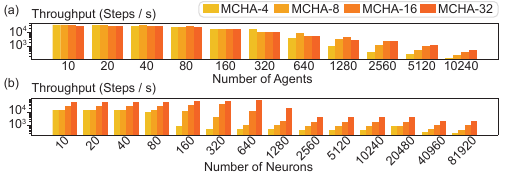}
    \caption{
        The scale-up effect of (a) SMAX (b) MVC.
    }
    \label{fig:10}
\end{figure}

\subsubsection{Roofline Model Analysis}

\dj{Figure~\ref{fig:9} shows the roofline models for \codename-4 and A100.}
Unlike conventional GPUs, which are typically bounded by a single global memory bandwidth (as shown in Figure~\ref{fig:1}(a)), \codename exhibits a tiered execution profile characterized by 4 bandwidth limits.
While the hierarchy of these bandwidth limits is fixed by the hardware design, the actual throughput bottleneck is dynamically determined by its specific algorithmic dataflow.
For example, \textit{STORM} and \textit{Switch Riddle} workloads are constrained by the Intra-Block Communication bandwidth.
Though the limitation is slight, this suggests that their high-frequency agent interactions saturate the local communication bandwidth before reaching the MCC's compute limits.
The remaining workloads are limited by the MCC computing throughput, operating within the bandwidth headroom provided by \codename-4.
This divergence emphasizes the key role of the memory-centric nature of the proposed data-driven programming principles.
\uline{By strategically organizing tasks to maximize data residency, \codename minimizes high-latency transmissions and shifts the operational point toward the compute-bound region for higher throughput.}

\subsubsection{Scaling-Up Effect}\label{sec:scale}

The architectural scalability of \codename is fundamentally driven by its high spatial and temporal locality.
Workloads with spatially organized neighborhood definitions, such as large-scale maps containing thousands of agents, are partitioned using a dynamic border strategy, as illustrated in Figure~\ref{fig:6}(e).
This allows the workload to be distributed across MCCs while maintaining data residency.
Conversely, for graph-based neighborhood topologies, a min-cut partitioning algorithm can be implemented to decompose the global graph into localized sub-graphs.
In this configuration, inter-sub-graph edge data is temporarily buffered in DRAM, ensuring that the high-frequency internal processing remains localized within the on-chip mesh.
Figure~\ref{fig:10} presents the scaling-up characteristics of the \codename hardware across varying workload sizes and hardware scales, using SMAX and MVC as the benchmark.
The persistent parameters and corresponding entities are spread evenly on the hardware for specific comparison.

Figure~\ref{fig:10}(a) shows that at small agent counts (10 to 40 agents in SMAX), the system throughput remains relatively constant.
This phenomenon occurs because the primary overhead of the execution at this time is not the compute intensity or communication volume, but the pipeline latency required to complete the initial dataflow sequence.
As the agent population grows, the dominant bottleneck shifts from pipeline latency to compute and transmission throughput.
In this regime, the processing time increases proportionally with the number of agents.
Additionally, Figure~\ref{fig:10}(a) also shows that the accelerating rates of \codename-32 compared with \codename-4 vary when the benchmark scale is different.
This is because the even distribution of the agents in the hardware leads to performance limits originating from the Inter-Chip Communication.
Similar trends can be observed from Figure~\ref{fig:10}(b) for the MVC benchmark.
Comparing Figure~\ref{fig:10}(a) with (b), for 2\% MVC benchmark, \codename-32 achieves 7.47$\times$ speedup compared with \codename-4, while this ratio decreases to 3.85$\times$ for SMAX, including 10240 agents.
The reason for this phenomenon is that the arithmetic density of SMAX is higher than that of the MVC with a reinforcement neural network included as the updating function.

\subsubsection{Design Space Exploration}

\begin{figure}[tb]
	\centering
	\includegraphics[width=1\columnwidth]{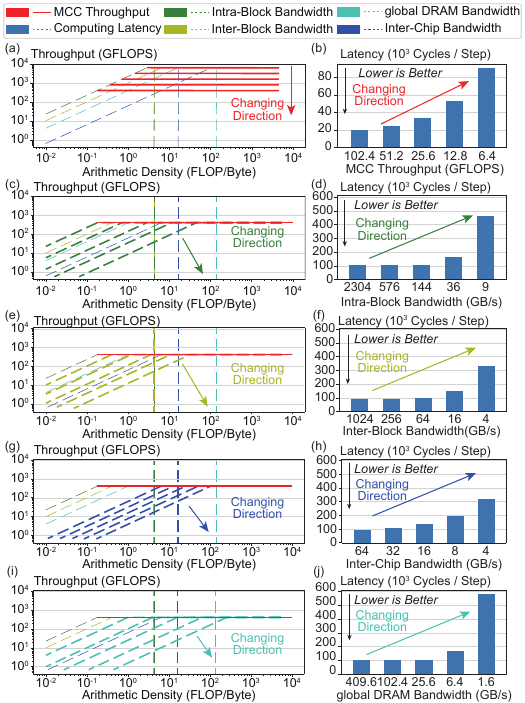}
	\caption{
		\dj{The roofline model and throughput comparison based on different (a--b) MCC throughput; (c--d) Intra-block bandwidth; (e--f) Inter-block bandwidth; (g--h) Inter-chip bandwidth; (i--j) global DRAM bandwidth.}
	}
	\label{fig:11}
\end{figure}

According to the roofline model presented in Section~\ref{sec:model}, the design space of \codename can be described by multiple bandwidths and computing limits.
Figure~\ref{fig:11} shows the influence of varying parameters in \codename-4 on the roofline models.
The varying design parameters cover MCC throughput (a--b) and the bandwidths of Intra-block (c--d), Inter-block (e--f), Inter-chip (g--h), and global DRAM (i--j).
Inter-chip bandwidth and global DRAM bandwidth have the largest impact on the overall effective throughput.
This phenomenon can be observed through the slight growth of the processing time before arriving at the turning point of the roofline model, as the decrease of inter-chip bandwidth and global DRAM bandwidth.
This indicates that the communication between different chips or off-chip memory should be controlled in this \codename architecture, which is straightforward for our design.
In other cases, the influence of the parameters on the final throughput is similar to the conventional roofline model.

\subsection{Related Works}

\codename vs. \textbf{Compute-Centric Architectures and Frameworks}.
While many-core CPUs and GPUs are the current mainstream hardware solution executing PSC tasks, they are doomed to hit the ``memory wall''~\cite{asanovic2006landscape, rogers2009scaling, ivanov2021data}.
Software frameworks like JAX-MARL~\cite{flair2024jaxmarl} attempt to alleviate this through JIT-compiled kernel fusion, yet they are still throttled by rigid hardware cache hierarchies.
For MARL workloads, this leads to massive cache coherence overhead, which can consume up to 96\% of total operations as shown in Figure~\ref{fig:1}(b).
In contrast, \codename abandons GBs, instead granting programmers explicit control over a reconfigurable memory fabric through the multi-tiered NoC.

\codename vs. \textbf{Near-Data Processing}.
While NDP and Processing-In-Memory (PIM) designs like Tesseract~\cite{ahn2015scalable} and UPMEM~\cite{friesel2023full} successfully reduce data movement, they are generally tailored for static irregularity, such as PageRank~\cite{page1999pagerank} or BFS~\cite{murphy2010introducing} on static graphs.
Conversely, PSC is defined by dynamic irregularity with an additional time dimension, where neighbor relationships change as the algorithm progresses.
Existing PIM implementations, such as StreamPIM~\cite{an2024stream}, frequently lack the inter-core bandwidth and architectural flexibility needed for the complex phase transitions of these tasks.
\codename bridges this gap by utilizing a multi-layer communication strategy for P2P entity interaction that releases the burden of off-chip memory access.

\codename vs. \textbf{Specialized Accelerators}.
Accelerators for specific irregular primitives can achieve high efficiency for their targeted kernels, such as pointer-chasing in graph processing (Minnow~\cite{zhang2018minnow}, GraphR~\cite{song2018graphr}) or sparse-tensor algebra (ExTensor~\cite{hegde2019extensor}).
However, they often lack support for general PSC tasks, making commercialization challenging with only supporting highly niche computing applications.
Furthermore, many DSAs still rely on a centralized global buffer, which becomes a serializing bottleneck as the computing scale increases.
\codename differentiates itself by proposing the data-driven programming model, enabling a reconfigurable fabric to support a diverse suite of workloads.

\section{Conclusion}
\dj{We propose \codename architecture, which accelerates PSC workloads, a critical paradigm spanning MARL, SNN, and BSP, by enabling P2P communication to relieve the global bandwidth pressure that bottlenecks conventional systems.} A data-driven programming model is provided for algorithm deployment. The architecture demonstrates strong data and processing locality, achieving speedups of 153.06$\times$ to 2456.96$\times$ over A100 GPUs in MARL tasks using a 4-chip system. Its generality and near-linear scalability are further validated through multi-application evaluation and roofline analysis. \dj{Two limitations merit mention: (a) \codename exploits locality through \textit{fixed} topologies, leaving highly dynamic graphs unsupported; (b) porting CPU/GPU code to \codename's programming model currently requires manual effort. We envision that advances in workload abstraction and agentic-AI-driven code-conversion toolkits will together unlock \codename's full potential.}

\bibliographystyle{unsrtnat}
\bibliography{ref}

\end{document}